\newcommand\pb{PbNi$_2$V$_2$O$_8$}
\newcommand\uu{\"{u}}
\newcommand\oo{\~{o}}
\newcommand\sr{SrNi$_2$V$_2$O$_8$}
\newcommand\PbNiMgVO%
{Pb(Ni${}_{1-x}$Mg${}_x$)${}_2$V${}_2$O$_8$}
\documentstyle[prb,aps,twocolumn]{revtex}
\begin{document}
\input{psfig.sty}
\draft

\twocolumn[\hsize\textwidth\columnwidth\hsize\csname
@twocolumnfalse\endcsname

\title{Zone-boundary excitations in coupled Haldane spin chain compounds \pb\ and \sr.}

\author{A. Zheludev}
\address{Physics Department, Brookhaven National Laboratory, Upton,
New York 11973 USA}

\author{T. Masuda and K. Uchinokura}
\address{Department of Advanced Materials Science and Department of Applied
Physics, The University of Tokyo,  Tokyo 113-8656, Japan.}

\author{S. E. Nagler}
\address{Oak Ridge National Laboratory, Bld. 7692, MS 6393, P.O. Box 2008, Oak Ridge, TN
37831, USA.}

\date{\today}
\maketitle
\begin{abstract}
Magnetic excitations in the quasi-one-dimensional quantum
antiferromagnets \pb\ and \sr\ are measured all the way up to the
zone boundary energy using inelastic neutron scattering from
powder samples. An estimate for next-nearest-neighbor in-chain
interactions is obtained. The role played by these interactions in
spin-vacancy induced magnetic ordering in \pb\ is discussed.
\end{abstract}


]

\section{Introduction}
The quasi-one-dimensional (quasi-1D) $S=1$ antiferromagnet (AF)
\pb\ was recently shown to be the first example of a Haldane-gap
system\cite{Haldane} to undergo long-range magnetic ordering upon
doping with non-magnetic impurities.\cite{Uchiyama99,Uchinokura00}
This unique behavior was understood through inelastic neutron
scattering measurements, that suggested that three dimensional
(3D) inter-chain interactions in \pb\ are almost (but not quite)
strong enough to destroy the Haldane singlet ground state even in
the undoped material.\cite{Uchiyama99,Zheludev00PbSrNi} In fact,
in the very similar isostructural compound \sr, inter-chain
coupling exceeds the critical value, and, unlike the Pb-based
material, the undoped Sr-Ni vanadate orders magnetically at
$T=7$~K. \cite{Uchiyama99,Zheludev00PbSrNi}

The introduction of non-magnetic vacancies into gapped quantum
spin chains produces free $S=1/2$ degrees of freedom on either
side of each impurity
site.\cite{AffleckKennedy88,Hagiwara90,Glarum91,Mitra92} In the
presence of arbitrarily weak inter-chain interactions, these free
spins will order in three dimensions at sufficiently low
temperature. However, next-nearest neighbor (nnn) interactions
within the chains can prevent this from happening.\cite{Shender91}
Naively, in the presence of AF nnn coupling, one can expect the
liberated spins around each impurity to bind into non-magnetic
singlets (dimers), preserving the disordered ground state of the
system.

There are at least two indications to that nnn interactions may be
active in \pb\ and \sr. First, the spin chains in both materials
are not linear, but are formed by Ni$^{2+}$ ions arranged in
4-step {\it spirals} propagating along the $c$ axis of the
orthorhombic lattice (for a detailed discussion of chain geometry
and inter-chain interactions see
Ref.~\onlinecite{Zheludev00PbSrNi}). In this ``twisted'' structure
nnn superexchange pathways involving equatorial oxygens of the
NiO$_6$ tetrahedra can be quite effective. Second, there remains
some discrepancy between the in-chain exchange constant $J$
deduced from the Haldane gaps measured with neutron scattering,
assuming only nearest-neighbor (nn) interactions ($J=9.0$~meV),
and from high-temperature ($T\approx J$) susceptibility data
($J=8.2$~meV). The former technique probes low-energy excitations
around $q_\|=\pi$ (AF zone-center), while the latter is sensitive
to excitations in the entire Brillouin zone. The mismatch may be
interpreted in terms of an additional coupling with a
characteristic interaction length of twice the spin-spin
separation in the chains, that is known to have different effects
on zone-center and zone-boundary excitation energies.

The most direct way to resolve this issue would be to measure the
dispersion of magnetic excitations in the entire Brillouin zone in
a single crystal sample. Unfortunately, no single-crystal or
aligned powder samples of either \pb\ or \sr, suitable for
inelastic neutron scattering studies, are available to date. As
will be discussed below, some information on nnn coupling can be
obtained using powder samples, if the measurements are extended
all the way to the zone-boundary energy. Previous neutron
scattering studies employed cold neutrons to investigate
low-energy excitations in the two
compounds.\cite{Uchiyama99,Zheludev00PbSrNi} In the present paper
we report thermal-neutron studies specifically aimed at measuring
high-energy excitations. We then engage in a semi-quantitative
discussion of the role nnn interactions play in vacancy-induced
ordering of weakly-coupled Haldane spin chains, and apply the
results to the particular case of \pb.
\section{Experimental procedures and results}
Both vanadates crystallize in  tetragonal structure, space group
$I41 cd$, with lattice constants $a = 12.249(3)$~\AA, $c =
8.354(2)$~\AA\ for \pb,\cite{Uchiyamaunpublished} and $a =
12.1617$~\AA, $c = 8.1617$~\AA\ for \sr,\cite{Wickman86}
respectively. All measurements were performed on large ($\approx
5$~g) powder samples of \pb\ and \sr, similar to those used in
previous cold-neutron studies.\cite{Zheludev00PbSrNi} The
experiments were carried out at the HB3 3-axis spectrometer
installed at the High Flux Isotope Reactor at Oak Ridge National
Laboratory. Neutrons with a fixed final energy of 13.5~meV were
used with a 5~cm thick Pyrolitic Graphite (PG) filter positioned
after the sample to eliminate higher-order beam contamination.
PG(002) reflections were used for monochromator and analyzer, in
combination with $48'-40'-40'-120'$ collimators. Sample
environment was in all cases a standard  He-4-flow cryostat. The
data were taken at $T=1.5$~K.
\begin{figure}
 \psfig{file=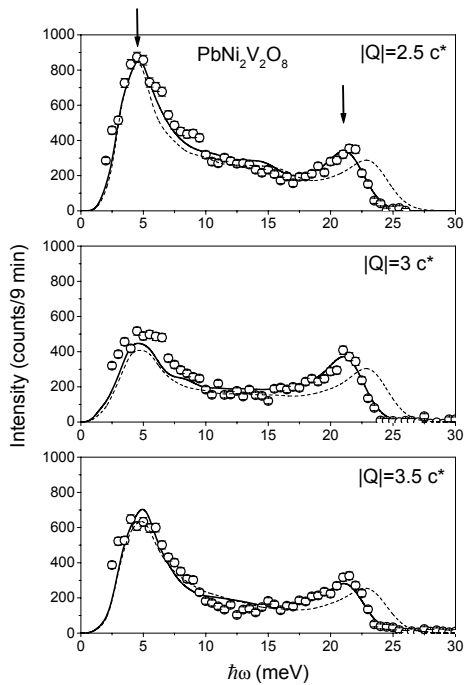,width=2.8in,angle=0}
 \caption{ Typical
background-corrected constant-$Q$ scans measured in \pb\ at
$T=1.5$~K. Arrows roughly indicate the gap and zone-boundary
energies. Solid lines are simulations based on the result of a
semi-global fir to the data, that allows for next-nearest-neighbor
in-chain interactions, as described in the text. Dashed lines are
a best fit assuming only nearest-neighbor interactions are
present. Statistical error bars are smaller than symbol size.
}\label{exdatapb}
\end{figure}

\begin{figure}
  \psfig{file=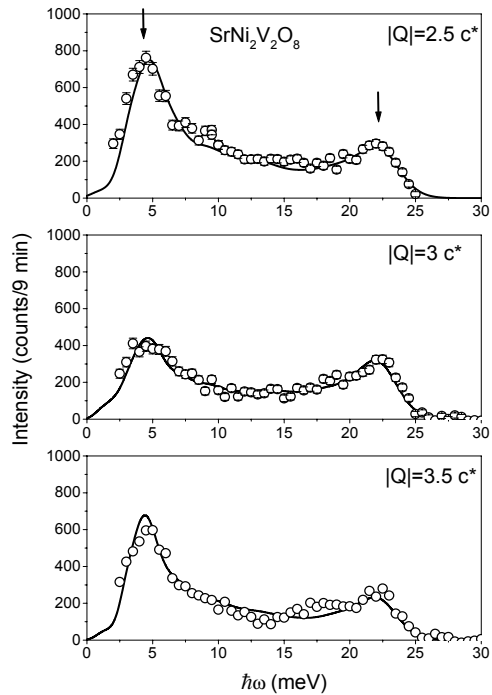,width=2.8in,angle=0}
 \caption{ Typical
background-corrected constant-$Q$ scans measured in \sr\ at
$T=1.5$~K. Arrows and solid lines are as in
Fig.~\protect\ref{exdatapb}. Statistical error bars are smaller
than symbol size.}\label{exdatasr}
\end{figure}

Typical data sets collected in constant-$Q$ scans for \pb\ and
\sr\ are shown in Fig.~\ref{exdatapb} and Fig.~\ref{exdatasr},
respectively. Figure~\ref{allsrdata} shows the bulk of the data
collected for the Sr system (only 5 scans, at $|Q|/c^{\ast}=$ 2.5,
3, 3.5, 4 and 5, were collected for the Pb-based material). The
measured intensities were scaled to correct for the previously
determined  energy-dependent $\lambda/2$ contribution to the
incident beam, that affects the monitor rate. Within the incident
neutron energy range used, the monitor efficiency changes by
roughly 30\%. In addition, a flat background (17 counts/min),
measured at 30~meV energy transfer (above the magnetic excitation
band), was subtracted from the data.

For both materials all scans have the characteristic M-shape. In a
powder sample one measures the spherical average of the dynamic
structure factor $S(\bbox{Q},\omega)$. For large momentum
transfers, a large portion of $\bbox{Q}$ space is sampled
simultaneously, and the powder cross section is in many ways
similar to the density of states function $n(\omega)$. In
particular, one expects to see a peak in the powder cross section
wherever the density of states is large, {\it i.e.}, at the gap
energy and at the zone boundary. As discussed
previously,\cite{Zheludev00PbSrNi} the peak around 5~meV energy
transfer corresponds to excitations close to the Haldane gap at a
momentum transfer $q_{\|}=\pi+2\pi n$ ($n$-integer) along the
chain axis. The weaker higher-energy peaks, represent excitation
near the top of the band, i.e., in the vicinity of
$q_{\|}=\pi/2+\pi n$ ($n$-integer).

\section{Data analysis}
\subsection{Model cross section}
The convoluted nature of inelastic powder data make it notoriously
difficult to interpret (for an example see
Ref.~\onlinecite{Zheludev96BaCuGeO}, particularly Section IV). The
only way to extract information on the dynamic structure factor
from such measurements is to assume a ``reasonable'' model and
refine the parameter values to best-fit the data. The main
assumption of the model used for \pb\ and \sr\ to date is that
these compounds are described as weakly-interacting $S=1$ AF
chains with magnetic anisotropy. This picture is based on
susceptibility, high field and specific heat measurements, and is
supported by the observed unique shape of constant-$E$ neutron
scans.\cite{Uchiyama99} In the present work we shall adopt this
model as well.

The form of the neutron scattering cross section for this model
was derived in Ref.~\onlinecite{Zheludev00PbSrNi}, and is written
in the single mode approximation (SMA). Inter-chain interactions
were accounted for within the Random Phase Approximation (RPA).
This cross section function performed rather well in describing
{\it low-energy} neutron data. Previous studies of other
Haldane-gap compounds, such as NENP\cite{Ma92} and
CsNiCl$_3$~\cite{Zaliznyak00}, have demonstrated that the SMA is,
in fact, applicable in most of the Brillouin zone. The previously
derived  SMA cross section (Eqs. 5--13 of
Ref.~\onlinecite{Zheludev00PbSrNi}) could thus be safely used in
interpreting our high-energy results. However, the expression for
the dispersion relation (Eqs.~14 and 15 of
Ref.~\onlinecite{Zheludev00PbSrNi}) had to be slightly modified.
These modifications are discussed in the following paragraphs.

\subsubsection{Isolated chains with nearest-neighbor interactions}
In Ref.~\onlinecite{Zheludev00PbSrNi} the dispersion of
excitations in non-interacting Haldane spin chains was assumed to
be sinusoidal and symmetric around the  classical AF zone-boundary
$q_{\|}=\pi/2$. In fact, since translational symmetry is not
broken in the Haldane ground state, the excitation energy
extrapolates to a larger value at $q_\|\rightarrow 0$ than at
$q_\|\rightarrow \pi$.\cite{Takahashi} This effect can be taken
into account empirically by
 including an extra term in the dispersion
relation.\cite{Ma92} The dynamic susceptibility of a single
Haldane spin chain is then written as:
\begin{eqnarray}
 \chi^{\beta\beta}(q_\|,\omega)=\chi^{\beta\beta}_\pi\frac{1-\cos
 q_\|}{2}
 \frac{\Delta_{\beta\beta}^2}{(\hbar\omega^{\beta\beta}_{q_\|})^2-(\hbar
 \omega +i \epsilon)^2},\\
 (\hbar\omega^{\beta\beta}_{q_\|})^2=\Delta_{\beta\beta}^2+v^2\sin^2q_\|+\alpha^2\cos^2\frac{q_\|}{2}.\label{disp}
\end{eqnarray}
In these formulas, which replace Eq.~22 in
Ref.~\onlinecite{Zheludev00PbSrNi}, $\Delta_{\beta\beta}$ is the
gap energy for the Haldane gap mode of a particular polarization
$\beta\beta$, $\chi^{\beta\beta}_\pi$ is the static staggered
susceptibility, and $v$ is the spin wave velocity. The spin chains
\pb\ and \sr\ contain four equivalent spins per crystallographic
period along the $c$ axis. Throughout this work the actual wave
vector transfer will be denoted as
$\bbox{Q}=h\bbox{a}^{\ast}+k\bbox{b}^{\ast}+l\bbox{c}^{\ast}$. The
symbol $q_\|\equiv \bbox{Qc}/4$ in Eq.~\ref{disp} will stand for
the reduced wave vector transfer along the chains.

The energetics of a Haldane-gap AF is by now very well established
through numerical work by several authors.\cite{num,num2,num3} For
an $S=1$ Heisenberg chain with single-ion anisotropy of type
$D(S^z)^2$ the approximate values of parameters are as follows:
\begin{eqnarray}
 \Delta_{xx}& = & \Delta_{yy}\approx \langle \Delta \rangle-\frac{2}{3}D,\label{e3}\\
 \Delta_{zz}& \approx & \langle \Delta \rangle+\frac{4}{3} D,\\
 \langle \Delta \rangle & \approx & 0.41 J,\label{e5}\\
 v & \approx &  2.49 J,\label{bind}\\
 \chi^{\beta\beta}_\pi & \approx & 1.26
 \frac{v}{\Delta_{\beta\beta}^2}.\label{e7}
\end{eqnarray}

The parameter $\alpha$ in Eq.~\ref{disp} characterizes the
asymmetry of the dispersion relation. Existing experimental data
for NENP\cite{Ma92} and CsNiCl$_3$ \cite{Zaliznyak00} gives
$\alpha=1.45 J\approx 0.58 v$ and $\alpha=1.1 J\approx 0.44 v $,
respectively. The latter results inspires more confidence, as it
was obtained with a much higher experimental resolution and gives
a dispersion relation very similar to that obtained by direct
diagonalization in finite-length chains.\cite{Takahashi}
Throughout the rest of the paper we shall assume $\alpha$ fixed at
$\alpha\equiv 0.44 v$.

\subsubsection{Next-nearest-neighbor interactions}
Next-nearest-neighbor in-chain interactions, that we believe can
be relevant for \pb\ and \sr, can be accounted for within the RPA.
Isotropic nnn coupling of magnitude $J'$  gives an additional term
in the r.h.s. of Eq.~\ref{disp} that can be written as:
\begin{equation}
 \chi^{\beta\beta}_\pi \Delta_{\beta\beta}^2 J'\case{1-\cos q_\|}{2}\cos2q_\|.\label{Jprime}
\end{equation}
The primary effect of nnn interactions is to modify the gap
energies:
\begin{eqnarray}
\Delta_{\beta\beta}^2=\Delta_{\beta\beta,0}^2\left(1+\chi^{\beta\beta}_\pi
J'\right)\label{dprime},
\end{eqnarray}
where $\Delta_{\beta\beta,0}$ are the gaps in the absence of nnn
coupling. The relative effect of nn interactions on the
zone-boundary energy and spin wave velocity is considerably
smaller, and can be ignored. To a good approximation, in analyzing
the experimental data one can use the dispersion
relation~\ref{disp}, but treat the gap energies and $v$ as
independent parameters, rather than assuming the
relation~\ref{bind} valid only in the case $J'=0$. The ratio
$\xi=v/\langle \Delta\rangle$ can then be determined  and used to
obtain an estimate for $J'$. From
Eqs.~\ref{e3}--\ref{e7},\ref{dprime} for $J'\ll J$ one gets:
\begin{equation}
\xi \approx \xi_0 (1-9.33 \frac{J'}{J})\label{estimate}
\end{equation}
Here $\xi_0=v/\langle \Delta_0 \rangle\approx 6.07$ is equal to
the equal-time correlation length in the absence of nnn coupling.

 \twocolumn[\hsize\textwidth\columnwidth\hsize\csname
 @twocolumnfalse\endcsname
\begin{figure}
 \psfig{file=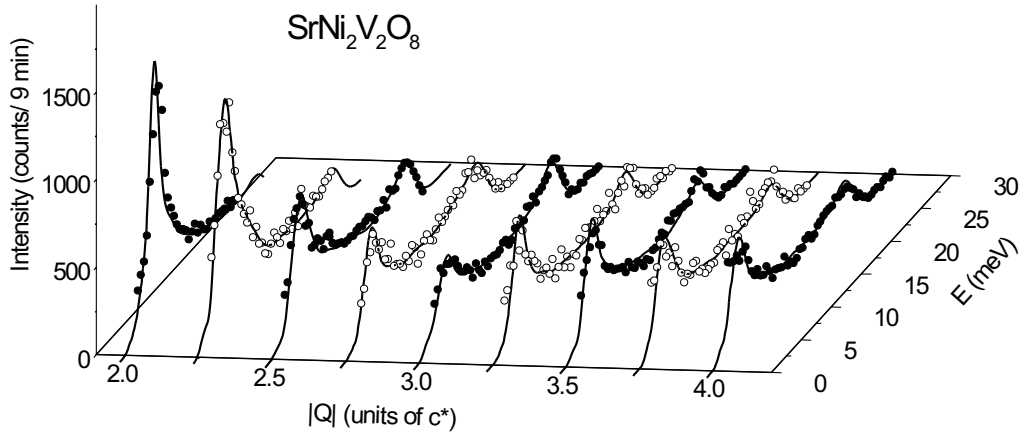,width=6in,angle=0}
 \caption{ The bulk of
constant-$Q$ scans measured in \sr\ at $T=1.5$~K. Solid lines are
as in Fig.~\protect\ref{exdatapb}. Statistical error bars are
smaller than symbol size.}\label{allsrdata}
\end{figure}
]

\subsubsection{Inter-chain interactions}
For the coupling geometry found in \pb\ and \sr, switching on
inter-chain interactions within the RPA gives an additional term
\begin{equation}
-\chi^{\beta\beta}_\pi\Delta_{\beta\beta}^2J_1\cos(\pi l/2)\left[
\cos(\pi h) +\cos(\pi k)\right]\frac{(1-\cos \pi
 l/2)}{2}\label{tran}
\end{equation} in the r.h.s. of Eq.~\ref{disp}. In this formula,
which corresponds to the last terms in Eqs.~14 and 15 of
Ref.~\onlinecite{Zheludev00PbSrNi}, $J_1$ is the magnitude of
nearest-neighbor inter-chain interactions. For our present
purposes, the possible small anisotropy of $J_1$ will be ignored.

\subsection{Parameters for \pb\ and \sr} Our understanding of the
physics of \pb\ and \sr\ is based on the knowledge of the
following characteristics: the intrinsic gap energies
$\Delta_{zz}$ and $\Delta_{xx}=\Delta_{yy}$, the magnitude of
inter-chain interactions $J_1$ and the spin wave velocity $v$. Can
these four parameters be unambiguously determined from the
available experimental data?

First, consider those quantities that characterize low-energy
excitations near $q_{\|}=\pi$. For \pb\ the actual minima of the
3D dispersion, $E_{{\rm min},\beta\beta}$ were {\it independently}
measured in high-field magnetization studies of {\it aligned}
powder samples.\cite{Uchiyama99,Zheludev00PbSrNi} These energies
are directly related to $\Delta_{\beta\beta}$ and $J_1$. The third
experimental value need to unambiguously ``untangle'' these three
quantities was probed in previous high-resolution cold neutron
measurements.\cite{Zheludev00PbSrNi} Indeed, the well-defined
lower-energy 5~meV peak in constant-$q$ powder scans is located
approximately at the top of the dispersion band perpendicular to
the chains at $q_\|=\pi$, and its position is directly related to
$\Delta_{\beta\beta}$ and $J_1$. From the combined high-field and
neutron measurements $\Delta_{\beta\beta}$ and $J_1$ can thus be
reliably determined within the established model. For \pb\ these
parameters are: $\Delta_{zz}=3.1(3)$~meV,
$\Delta_{xx}=\Delta_{yy}=4.0(3)$~meV and
$J_1=-0.17(2)$~meV.\cite{Zheludev00PbSrNi}

For \sr, which is {\it ordered} at low temperature, the model of
weakly-coupled Haldane spin chains is, strictly speaking, not
applicable. Nevertheless, as the ordered moment is expected to be
very small, the model is still a good approximation, except in the
direct proximity of the magnetic Bragg peaks. From the analysis of
the low-energy powder spectrum for \sr\ the following parameters
were previously determined: $\Delta_{zz}=2.8(4)$~meV,
$\Delta_{xx}=\Delta_{yy}=3.9(3)$~meV and
$J_1=-0.17(3)$~meV.\cite{Zheludev00PbSrNi}
\begin{table}
\caption{Parameters of the model Hamiltonian used to analyze the
experimental data.}
\begin{tabular}{lcc}
 Parameter& \pb & \sr \\
 \tableline
 $\Delta_{zz}$ & 3.1 (fixed)& 2.8 (fixed)\\
 $\Delta_{xx}=\Delta_{yy}$ & 4.0 (fixed)& 3.9 (fixed)\\
 $J_1$ & -0.17 (fixed)& -0.17 (fixed)\\
 $\alpha$ & 1.1 (fixed)& 1.1 (fixed)\\
 $v$ & 20.6(7)&  21.2 (4)
\end{tabular}
\end{table}

The exact value of the remaining ``high-energy'' parameter $v$ has
little impact on the low-energy powder spectrum, and was not
treated as an independent variable in
Ref.~\onlinecite{Zheludev00PbSrNi}. In contrast, $v$ determines
the high-temperature magnetic susceptibility of the system, that,
provided $k_{\rm B}T \gg \Delta$, are insensitive to the details
of the low-energy spectrum. This type of experiments yields
$v\approx20.4$~meV for \pb. The present study provides the most
direct measure of the zone-boundary energy, and hence the spin
wave velocity $v$, defined by the location of the higher-energy
peak and cutoff in the constant-$q$ powder scans. In analyzing
these data we shall use the cross section defined in Eqs.~5--13 of
Ref.~\onlinecite{Zheludev00PbSrNi} in combination with the
dispersion relation given by Eqs.~\ref{disp} and \ref{tran} above,
using only $v$ as an adjustable parameter of the model.

\subsection{Fits to experimental data.}
In the data analysis, the powder-average of the parameterized
cross section was calculated numerically using a Monte Carlo
routine. The result was further numerically convoluted with the
$Q-E$ resolution function of the 3-axis spectrometer, calculated
in the Cooper-Nathans approximation.\cite{CooperNathans}
Performing a true global fit to the data, using a single set of
parameters to describe all measured scans, was not possible
because of the scattering-angle dependence of neutron transmission
in the sample. The effect is due to a large sample size and
appreciable incoherent cross sections of Ni and V nuclei.
Transmission in the sample was studied by measuring the intensity
of incoherent elastic scattering, and found to change by
$\approx30$\%, depending on scattering geometry. Within each
constant-$Q$ scan the correction is small and could  be ignored.
However, in analyzing the entire data sets, collected in a broad
range of momentum transfers (scattering angles), a separate
intensity prefactor had to be used for each const-$Q$ scan. Thus,
for each compound, the data were analyzed in a semi-global fit: a
single adjustable parameter $v$ was used to {\it simultaneously}
describe all measured scans, each with its own adjustable
prefactor. The parameters were refined to best-fit the data using
a least-squares algorithm.

Excellent fits were obtained with $v=20.6(7)$~meV for \pb\ and
$v=21.2(4)$~meV for \sr. Simulations based on the refined
parameter values are shown in solid lines in
Figs.~\ref{exdatapb}--\ref{allsrdata}. The fits correspond to
$\chi^2=4.7$ for \pb\, and $\chi^2=3.8$ for \sr. Given that the
overall shape of the measured scans is reproduced by our model
quite well, the large values of $\chi^2$ are most likely due to
the fact that experimental statistical error bars are less
significant than unavoidable systematic errors, such as
multi-phonon scattering in the sample. While such effects can lead
to an inaccurate background subtraction, they are not likely to
influence the measured zone-boundary energy, as the latter is
accurately pin-pointed by the well-defined higher-energy cutoff in
the energy scans. The relevant parameters used to fit the data,
both fixed and adjustable, are summarized in Table~\ref{table} for
both systems. The refined values for $v$ are to be compared to
previous estimates based on the measured gap energies and
Eq.\ref{e5}, assuming only nearest-neighbor interactions,
$v=22.5(1.0)$~meV and $v=21.5(0.9)$~meV, for \pb\ and \sr,
respectively.\cite{Zheludev00PbSrNi} For comparison, the dashed
lines in Figs.~\ref{exdatapb} are simulations based on these
latter value for \pb.

It was important to verify that any uncertainty that may exist for
the dispersion asymmetry parameter $\alpha$ does not have a
significant impact on our determination of $v$. This was done by
analyzing the \pb\ data using a larger value $\alpha=0.58v$, as
previously experimentally determined for NENP,\cite{Ma92} rather
than $\alpha=0.44v$. Essentially the same quality of fits was
obtained, and the refined spin wave velocity $v=19.9(8)$~meV is
within the error bar of that obtained with $\alpha=0.44v$, as
described above. This demonstrates that the data analysis
procedure employed is fairly insensitive to dispersion asymmetry.

We see that any deviations from the model involving only
nearest-neighbor interactions are small in either material. While
for \sr,  within the error bars, the discrepancy is undetectable,
for \pb\ it appears to be more substantial. It is gratifying to
see that for the latter system, the bandwidth measured in this
work is in excellent agreement with estimates based on the
high-temperature susceptibility (see above). As explained in the
previous section, the reduction of zone-boundary energy is a sign
of antiferromagnetic nnn in-chain interactions. Using
Eq.~\ref{estimate} for \pb\ we obtain $J'\approx0.4(2)$~meV.
Detecting such small values of $J'$ is only possible due to the
large staggered susceptibility of a Haldane spin chain (the gap is
relatively small compared to $v$), and hence the large prefactor
in Eq.~\ref{estimate}.

\section{Discussion}
To summarize, the previously proposed nearest-neighbor model for
\pb\ and \sr\  should be an adequate description of these
compounds in most cases, in the entire Brillouin zone. For \pb\
the data suggest, barely outside the error bars, a small
antiferromagnetic nnn in-chain coupling of a few tenths of a meV.
But how strong should this coupling be to have an impact on
long-range ordering in a doped material? We shall now discuss the
general significance of nnn interactions to spin-vacancy-induced
order in weakly-coupled integer-spin quantum chains, and then
consider the particular case of \pb.

\subsection{Interactions between end-chain spins}
The problem of spin-vacancy-induced ordering in coupled
integer-spin quantum chains with {\it only} nearest-neighbor
in-chain interactions has been studied theoretically by Shender
and Kivelson.\cite{Shender91}  The introduction of a non-magnetic
impurity in place of a $S=1$ ion  produces two free $S=1/2$ spins
located on the two ends of the severed
chain.\cite{AffleckKennedy88,Hagiwara90,Glarum91,Mitra92} In each
of the two fragments the end-chain spin is delocalized on a length
scale defined by the intrinsic equal-time correlation length
$\xi\approx 6$ of a Haldane spin chain. In principle, existing 3D
interactions in the system should lead to long-range magnetic
ordering of these liberated end-chain spins at a non-zero
temperature.

\begin{figure}
 \psfig{file=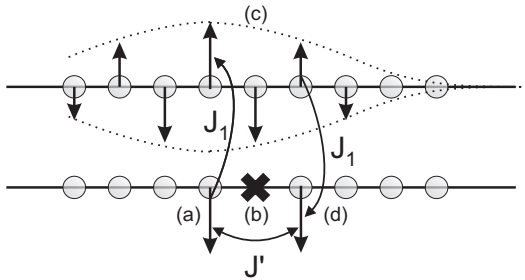,width=3.2in,angle=0}
 \caption{ A schematic
representation of magnetic interactions around a non-magnetic
impurity site. End-chain $S=1/2$ degrees of freedom (a and d)
appear to either side of the impurity site (b). For simplicity,
the delocalized nature of these end-chain spins is not shown. The
interaction $J_1$ of a liberated spin (a) with a neighboring chain
locally polarizes the latter, inducing a ``pocket'' of staggered
magnetization (c). The induced magnetization, in turn, couples to
the other free spin (d). The resulting effective coupling competes
with direct next-nearest-neighbor in-chain interactions $J'$.
 }\label{coupling}
\end{figure}

The most interesting result of Ref.~\onlinecite{Shender91} is that
if inter-chain interactions are below the threshold value at which
even the undoped system orders in three dimensions, there must be
a critical spin-vacancy concentration $x_c$ below which the system
remains disordered at $T=0$. The mechanism of this behavior is
illustrated in Fig.~\ref{coupling}, which show spins in the
vicinity of one non-magnetic impurity. For simplicity, the
delocalized nature of end-chain spins is not shown. By virtue of
inter-chain coupling $J_1$ the end-chain spin on one of the
fragments of the severed chain induces a ``pocket'' of staggered
magnetization on a neighboring chain. The size of the pocket is
determined by the correlation length $\xi$. The induced staggered
moment interacts (again via $J_1$) with the end-chain spin on the
other fragment. This adds up to an effective 2nd-order coupling
between end-chain spins of the order of $|J'_{\rm eff}|\approx
z\chi_{\pi}J_1^2$, where $z$ is the inter-chain coordination
number. In Ref.~\onlinecite{Shender91} it is argued that this
coupling is antiferromagnetic and will bind the two end-chain
spins into a non-magnetic singlet. As a result, the system will
remain in a singlet ground state despite the introduction of a
vacancy. Another way of looking at this phenomenon is that the
effective nnn interaction $J'_{\rm eff}$ ``patches'' the broken
Haldane chain at the impurity site with an antiferromagnetic bond,
albeit a weak one.

We believe that this argument is, in general, incorrect. Indeed,
from Fig.~\ref{coupling} it is clear that, no matter what the sign
of $J_1$ is, $J'_{\rm eff}$ is actually {\it ferromagnetic}
($J'_{\rm eff}<0$) and, if anything, binds the two end-chain spins
into a {\it magnetic} triplet, delocalized over the length scale
$\xi$. Thus, in the absence of nnn interactions, long-range
ordering should occur for arbitrary small impurity concentration
$x$. The ordering temperature will, of course, be exponentially
small with $x$.\cite{Shender91}

This conclusion has to be modified in the presence of actual
in-chain nnn AF interactions. For $J'\gtrsim|J'_{eff}|$ the
arguments of Ref.~\onlinecite{Shender91} can be repeated verbatim
to show that the system will remain disordered at $T=0$ for
sufficiently small $x$. On the other hand, the doped system will
order at a non-zero temperature for $J'\lesssim |J'_{eff}|$. At
any given small concentration there will be a quantum phase
transition from a disordered to ordered state at a critical
strength of $J'$ given by
 \begin{equation}
 J'_c\approx z\chi_\pi J_1^2.\label{crit}
 \end{equation}
These estimates are semi-quantitative at best, yet they provide a
simple physical picture and emphasize the relevant energy scales.

\subsection{Application to \pb}
In \pb, that remains in a gapped state at $T\rightarrow 0$,
long-range ordering can be induced by substituting some of the
$S=1$ Ni$^{2+}$ ions with non-magnetic Mg$^{2+}$
ions.\cite{Uchiyama99} The transition was observed for all
Mg-concentrations $x$ ranging from 0.1 down to
0.01.\cite{Uchinokura00} Though the Neel temperature $T_{\rm N}$
was found to decrease rapidly with decreasing $x$, there is, to
date, no clear evidence of a critical concentration $x_c$ below
which the system remains disordered. The fact that in \pb\
doping-induced ordering is related to end-chain spins has been
elegantly demonstrated by comparing the effects of Mg$^{2+}$
($S=0$) and Cu$^{2+}$ ($S=1/2$) substitution on the Ni-sites. The
latter is exactly half as effective, in terms of the effect on
magnetic susceptibility, since the Cu-spin and the two end-chain
spins become bound in a doublet state. In contrast with $S=0$
impurity substitution, the introduction of a $S=1/2$ impurity
produces only one, not two, free spins in the
system.\cite{Uchiyama99}

To estimate the critical strength of nnn AF interactions that
would interfere with long-range ordering in weakly-doped \pb, we
have to estimate $J'_{\rm eff}$. Using the experimental values of
$J$, and assuming $z=2$,\cite{Zheludev00PbSrNi} from
Eqs.~\ref{crit} and \ref{e5}--\ref{e7} we obtain $J_c=|J'_{\rm
eff}|= -0.18(6)$~meV. Given the experimental error bars and the
semi-quantitative nature of Eq.~\ref{crit}, we can only conclude
that in \pb\ $J'$ is of the same order of magnitude as the
critical value $J'_c$. The question of whether there is a critical
doping level below which \pb\ remains disordered, stays open.
Hopefully, extending the $T_{\rm N}(x)$ measurements to lower
concentrations, more accurate single-crystal neutron scattering
measurements of the gap energies, and the development of a
quantitative theoretical model will resolve this interesting issue
in the future.

\section{conclusions}
i) This work provides a direct and accurate measurement of the
zone-boundary energy in \pb\ and \sr. The previously proposed
model, based on weakly-interacting $S=1$ quantum spin chains with
nearest-neighbor interactions, is shown to be an adequate
description of these compounds in the entire dynamic range of
magnetic excitations. ii) Unlike thought previously,
weakly-coupled Haldane spin chains should order in three
dimensions at {\it arbitrary small} concentrations of non-magnetic
impurities. iii) Next-nearest neighbor AF interactions can inhibit
long-range ordering at small impurity concentrations, but only if
they exceed a certain critical value $J'_c$. iv) In the particular
case of \pb\ nnn interactions, if at all present, are no larger
that several tenths of a meV, and may be of the same order of
magnitude as $J'_c$.

\acknowledgements We would like to thank Y. Uchiyama and I.
Tsukada for the collaboration in the early stage of  this study,
A. Tsvelik for fruitful discussions, and B. Taylor, R. Roth and K.
Mohanty for technical support. Work at Brookhaven National
Laboratory was carried out under Contract No. DE-AC02-98CH10886,
Division of Material Science, U.S.\ Department of Energy. The work
at the University of Tokyo was partially supported by Grant-in-Aid
for COE Research ``SCP Project" from the Ministry of Education,
Science, Sports and Culture.  Oak Ridge National Laboratory is
managed for the U.S. D.O.E. by Lockheed Martin Energy Research
Corporation under contract DE-AC05-96OR22464.


\end{document}